\documentclass[twocolumn,aps,pre,floatfix,superscriptaddress]{revtex4-2}
\usepackage{amsthm,amsmath,amssymb,bbm,hyperref,graphicx,float,tikz,comment,mathrsfs}
\usepackage{bbold}
\usepackage{color}
\usepackage[dvipsnames]{xcolor}
\usepackage{titlesec}

\usepackage[inline]{enumitem} % enumerated lists
\usepackage{hyperref}
\usepackage{appendix}

\usepackage{epsfig}

\usepackage{pgfplots}
\pgfplotsset{compat=1.18}

\usepackage{wrapfig}

\usepackage{graphicx}
\usepackage{subcaption}

\usepackage{pgfplots}
\pgfplotsset{compat=newest}
\usepackage{xcolor}
\usepackage{graphicx}
\usepackage[left=23mm,right=13mm,top=35mm,columnsep=15pt]{geometry} 
\usepackage{adjustbox}
\usepackage{placeins}
\usepackage[T1]{fontenc}
\usepackage{lipsum}
\usepackage{csquotes}
\usepackage{dsfont}

\usepackage[normalem]{ulem}

\makeatletter
\renewcommand{\@makecaption}[2]{%
  \par\vskip\abovecaptionskip
  \begingroup
    \small\parindent\z@
    \leftskip\z@\@plus\z@ \rightskip\z@\@plus\z@
    \parfillskip\z@\@plus1fil
    #1: #2\par
  \endgroup
  \vskip\belowcaptionskip
}
\makeatother

\AtBeginDocument{%
}

\newtheorem{sta}{Statement}
\newtheorem{thm}{Theorem}
\theoremstyle{definition}
\newtheorem{rmk}{Remark}

\DeclareMathOperator{\tr}{tr}

\DeclareMathOperator{\vol}{vol}
\newcommand{\mc}{\mathrm{mc}}
\newcommand{\eq}{\mathrm{eq}}

\newcommand\norm[1]{\lVert#1\rVert}
\newcommand{\Hilbert}{\mathscr{H}}

\newcommand{\be}{\begin{equation}}
\newcommand{\ee}{\end{equation}}
\newcommand{\EEE}{\mathbb{E}}
\newcommand{\RRR}{\mathbb{R}}
\newcommand{\SSS}{\mathbb{S}}
\newcommand{\qB}{\mathrm{qB}}
\newcommand{\B}{\mathrm{B}}

\newcommand{\W}{|| P_\nu \psi_t||^2}

\newcommand{\PWig}{\mathbb{P}_\text{Wig}}

\begin{document}

\title{On the Increase of Boltzmann Entropy in Closed Macroscopic Quantum Systems}

\author{Cedric Igelspacher}
\email{E-mail: cedric.igelspacher@uni-tuebingen.de}
\affiliation{Mathematics Institute, Eberhard Karls University T\"ubingen, Auf der Morgenstelle 10, 72076 T\"ubingen, Germany.}

\author{Roderich Tumulka}
\email{E-mail: roderich.tumulka@uni-tuebingen.de}
\affiliation{Mathematics Institute, Eberhard Karls University T\"ubingen, Auf der Morgenstelle 10, 72076 T\"ubingen, Germany.}
\date{October 26, 2026}

\begin{abstract}
We consider a closed macroscopic quantum system in a pure state $\psi$ evolving unitarily with Hamiltonian $H$ in a finite-dimensional Hilbert space $\mathscr{H}$. Following von Neumann, we assume a decomposition of $\mathscr{H}$ into mutually orthogonal subspaces $\mathscr{H}_{\nu}$, where each $\nu$ represents a macro state, and $\mathscr{H}_{\nu}$ contains the pure states compatible with $\nu$. We define the quantum Boltzmann entropy for a macro state $\nu$ as $S_\qB(\nu) = k_{\mathrm{B}} \log \dim \mathscr{H}_{\nu}$, where $k_\B$ is the Boltzmann constant. Goldstein et al.\ (2019) \cite{GLTZ_19} have suggested, as a version of the second law of thermodynamics, that $S_\qB$ predominantly increases with time for realistic macroscopic systems. Here, we discuss precise versions of this statement and confirm them for several simple models of $H$ and $\mathscr{H}_\nu$.
\end{abstract}


\maketitle

\section{Introduction}

We are interested here in concrete formulations of the second law of thermodynamics, understood as the statement that entropy increases with time \cite{Cla65}, and derivations of such formulations from quantum mechanics. Specifically, we are concerned with definitions of entropy that represent, roughly speaking, the logarithm of the number of micro states compatible with a macro state (``Boltzmann entropy''), as opposed to definitions referring to the knowledge of observers or the spread of probability distributions. Correspondingly, the kind of statement we consider for the second law asserts that systems tend to evolve from one kind of micro state (regarded as having low entropy) to another kind of micro state (having high entropy), also in the absence of observers and for closed systems, and is not limited to transitions between thermal equilibrium states with different parameters. 

We therefore consider a closed macroscopic quantum system $\mathscr{S}$ confined to some bounded volume, say a gas in a box, and use the \emph{quantum Boltzmann entropy} 
\cite{Pen79,Leb_93,Gri_94,Leb_07,GLTZ_EPJH,GLTZ_19} (foreshadowed already in Einstein's work \cite{Ein_14})
\begin{equation}\label{eq:QBE}
    S_{\textrm{qB}}(\nu) = k_{\B} \log \dim \Hilbert_\nu,
\end{equation}
where $k_\B$ is the Boltzmann constant and $\Hilbert_\nu$ the ``macro space'' containing the pure states compatible with macro state $\nu$ (see Section~\ref{sec:BQF}).
Microscopically, the state of $\mathscr{S}$ is given by a wave function $\psi$ in the system Hilbert space $\Hilbert$ evolving unitarily, 
\begin{equation}\label{eq:unitary}
    \psi_t = e^{-iHt} \psi_0\,,
\end{equation}
with Hamiltonian $H$ (we set $\hbar=1$) and initial state $\psi_0\in\Hilbert$. The ``individualistic'' (as opposed to ``ensemblistic'') approach of studying thermodynamic behavior in terms of a single closed system in a pure state \cite{vNeu_29,GLTZ_EPJH,Gol_19} has become widespread in particular in connection with the eigenstate thermalization hypothesis and canonical typicality \cite{GE_16,Mori_18,BGVZ_25}.

Here, we propose two precise versions of the statement that $S_\qB$ increases with time, which we call the ``weak second law'' and the ``strong second law.'' It would be of interest to prove that they hold for a class of systems that covers realistic macroscopic systems. Here, we show that they hold for two simple model systems, and point out that they also hold for a third model recently analyzed by Erd\H os, Henheik, and Vogel \cite{EHV_25}. The models, outlined in Section~\ref{sec:model} and detailed in Sections~\ref{sec:ex1}--\ref{sec:ex3}, involve: first, only two macro spaces, of which one has low dimension; second, a random $H$ from the Gaussian unitary ensemble (GUE) or a Wigner ensemble; and third, a random block-band matrix as $H$.

The paper is organized as follows: we begin in Section~\ref{sec:setup} by outlining the framework, our definition of entropy, and the model systems considered. In Section~\ref{sec:2BQF}, we suggest two formulations of the second law. Afterwards in Sections~\ref{sec:ex1} and \ref{sec:ex2}, we confirm them for the first two models. In Section~\ref{sec:ex3}, we report what can be concluded about increase of $S_\qB$ for the third model. In Section~\ref{sec:outlook}, we provide a comparison to the literature and outlook. We collect the proofs and calculations in the Appendix.

\section{Setup}
\label{sec:setup}

\subsection{Boltzmannian Quantum Framework}\label{sec:BQF}

Let $\Hilbert_\mc\subseteq \Hilbert$ denote the micro-canonical energy shell, i.e., the Hilbert space spanned by the eigenvectors of $H$ with eigenvalues in $[E-\Delta E, E]$ with $\Delta E>0$ being macroscopically small, but still large enough to contain a huge number of eigenvalues of $H$; let $D$ be the dimension of $\Hilbert_\mc$. Following von Neumann \cite{vNeu_29}, we assume that all macroscopic observables $M_1,\ldots,M_K$ are highly degenerate (because macroscopic measurements cannot finely distinguish between different micro states) and commute with each other (because they can be jointly measured). As a description of the macro state $\nu$, we use the joint eigenvalues of the macro observables, $\nu=(m_1,\ldots,m_K)$; let the $\Hilbert_\nu$ be the joint eigenspaces of the $M_1,\ldots,M_K$. This provides a decomposition of $\Hilbert$ into mutually orthogonal subspaces (see also \cite{Leb_93,GLMTZ_10a,GLMTZ_10n,GLTZ_EPJH,TTV_23}),
\begin{equation}\label{eq:decomp}
    \Hilbert=\bigoplus_{\nu\in\mathscr{N}} \Hilbert_\nu 
\end{equation}
with suitable index set $\mathscr{N}$. Since coarse-grained energy should be among the macro observables, and since they commute, we can also assume that this decomposition aligns with the decomposition into energy shells, so
\begin{equation}\label{eq:decompmc}
    \Hilbert_\mc=\bigoplus_{\nu\in\mathscr{N}_\mc} \Hilbert_\nu
\end{equation}
for a suitable subset $\mathscr{N}_\mc \subseteq \mathscr{N}$. In the following, we will not talk further about the macro observables but regard the macro spaces $\Hilbert_\nu$ as given, with the properties \eqref{eq:decompmc} and that their dimensions $d_\nu:=\dim\Hilbert_\nu$ are very large; $P_\nu$ will denote the projector onto $\Hilbert_\nu$.

To each $\nu$ one can then assign the quantum Boltzmann entropy $S_\qB(\nu)$ as in \eqref{eq:QBE}. For systems $\mathscr{S}$ with $\psi$ lying in one of the $\Hilbert_\nu$, this also defines an entropy value of $\mathscr{S}$, but most $\psi$, of course, are non-trivial superpositions of contributions from several $\Hilbert_\nu$'s. Of them, one should think as being superpositions of different entropy values (just as they can be superpositions of different values of any quantum observable). Correspondingly, one can introduce the \emph{entropy operator} \cite{GLTZ_19}
\be
\hat{S}= \sum_{\nu \in \mathscr{N}} S_\qB(\nu) \, P_\nu \,.
\ee

$S_\qB$ has several properties expected of thermal entropy: (i)~It is additive when we consider a composite of two macroscopic systems, $\Hilbert=\Hilbert_a \otimes \Hilbert_b$, whose macro states $\nu=(\nu_a,\nu_b)$ consist of specifying the individual macro states of each system so that $\Hilbert_\nu=\Hilbert_{\nu_a} \otimes \Hilbert_{\nu_b}$ \cite{GLTZ_19}. (ii)~It parallels the classical definition of Boltzmann entropy,
\be
S_\mathrm{B}(\nu)=k_\mathrm{B} \, \log \vol \Gamma_\nu
\ee
with $\Gamma=\bigcup_\nu\Gamma_\nu$ the partition of phase space $\Gamma$ into ``macro sets'' $\Gamma_\nu$ containing all phase points compatible with $\nu$ \cite{Leb_07}. (iii)~For thermal equilibrium macro states, it yields the correct values as follows. In every $\Hilbert_\mc$ (or the corresponding subspace in the presence of macroscopic conserved observables \cite{ITV_26}), there is one macro state corresponding to thermal equilibrium, $\nu=\eq$, and the corresponding subspace $\Hilbert_\eq \subseteq \Hilbert_\mc$ has the overwhelming majority of dimensions,
\be\label{deqD1}
d_\eq/D \approx 1.
\ee
Of a system $\mathscr{S}$ for which $\psi\in\Hilbert_\eq$ (at least approximately), one says that it is in \emph{macroscopic thermal equilibrium} (MATE)  \cite{GLMTZ_10a,GHLT_15,GHLT_17}. As a consequence of \eqref{deqD1}, $S_\qB$ is maximal in thermal equilibrium, and in fact
\be
S_\qB(\eq)\approx k_\mathrm{B} \, \log D\,.
\ee
This value seems to be widely confirmed and accepted as correct, and it agrees with the von Neumann entropy $S_\mathrm{vN}(\rho)= -k_\mathrm{B} \tr(\rho \log \rho)$ for $\rho=\rho_\mc$; thus, the two definitions of entropy agree for systems in thermal equilibrium.

To say that entropy increases with time then means that the unitary evolution tends to carry states from smaller to larger subspaces, rather than the other way around. For superpositions of different entropy values, we still need to talk about what we mean by such a statement (see Section~\ref{sec:2BQF}), but at least such a statement sounds roughly plausible whenever $H$ is not chosen in a very special way tailored to the decomposition $\Hilbert=\bigoplus_\nu \Hilbert_\nu$. In fact, as is easy to see, $S_\qB$ increases whenever the time evolution is macroscopically deterministic: that is, if, for fixed $t,\nu_0,\nu$ and every $\psi_0\in\Hilbert_{\nu_0}$, $\psi_t$ lies in $\Hilbert_\nu$, then $d_\nu \geq d_{\nu_0}$; put differently, \cite{GLTZ_19}
\begin{equation}\label{eq:NatInc}
    \mathrm{if} \, \, e^{-iHt}\Hilbert_{\nu_0} \subseteq \Hilbert_{\nu}, \, \mathrm{then} \, \, S_\qB(\nu)\geq S_\qB(\nu_{0}).
\end{equation}

\subsection{Model Systems}\label{sec:model}

In Sections~\ref{sec:ex1} and \ref{sec:ex2}, we will consider two toy models of $H$ and $\Hilbert_\nu$, for which we can prove results about an increase of $S_\qB$ in the limit $D\to\infty$:

\begin{enumerate}
\item Assume that there are only two macro spaces, $\Hilbert_1$ and $\Hilbert_2$, that $d_1/D\ll 1$ and $d_2=D-d_1$, and that the direction of $\Hilbert_1$ is chosen purely randomly (i.e., uniformly) in $\Hilbert_\mc$. Of $H$, we assume that it is meaningful to speak of the \emph{density of states} (i.e., of a statistical distribution density). 
\item Assume, for any choice of the $\Hilbert_\nu$, that $H$ is a random matrix from the GUE (i.e., $H=A+A^\dagger$, where $\mathrm{Re}\, A_{ij}$ and $\mathrm{Im}\, A_{ij}$ of the entries $A_{ij}$ of the $D\times D$ matrix $A$ are independent Gaussian random variables with mean 0 and variance $1/4D$), or more generally a Wigner matrix (defined as before but without the requirement that the $\mathrm{Re} \, A_{ij}$ and $\mathrm{Im} \, A_{ij}$ are Gaussian, just independent and identically distributed with finite moments).
\end{enumerate}

It should be mentioned here that, although serving as simple case studies, GUE and Wigner matrices are physically unrealistic models of $H$ as they lead to ultrafast thermalization \cite{GHT_13b,Rei_16}. The same ultrafast thermalization occurs when the direction of $\Hilbert_\nu$ is purely random.

\begin{rmk}
Ultrafast thermalization arises from the fact that transitions from (more or less) any state directly into $\Hilbert_\eq$ are equally possible; this phenomenon also occurs classically when we model the phase space as a finite set $\Gamma$ and a discrete time evolution $T^n$ given by a permutation $T:\Gamma\to\Gamma$ in every time step: for any given initial phase point $X_0\in\Gamma$ and any given partition $\Gamma=\bigcup_\nu\Gamma_\nu$ such that a certain set $\Gamma_\nu=\Gamma_\eq$ contains the overwhelming majority of points, most permutations (or, equivalently, a uniformly distributed permutation) will carry $X_0$ to $\Gamma_\eq$ in the first time step.$\hfill \diamond$ \end{rmk}

For a third, somewhat more realistic model that avoids ultrafast thermalization, we will describe what follows about increase of $S_\qB$ from the analysis of \cite{EHV_25}:
\begin{enumerate}
\setcounter{enumi}{2}
\item Consider three macro spaces, $\Hilbert_\mc=\Hilbert_1 \oplus \Hilbert_2 \oplus \Hilbert_3$, and assume that, in an orthonormal basis aligned with the $\Hilbert_\nu$, $H$ is a random block-band matrix; that is, its entries are random with mean zero and variances that are constant in each of the nine blocks defined by the three subspaces and become smaller with the distance of the block from the diagonal. In fact, $H$ allows transitions only between \emph{neighboring} macro states.
\end{enumerate}

\section{Criteria for Increase of Boltzmann Entropy}\label{sec:2BQF}

In this section, we introduce two relevant conditions, the ``weak second law'' and the ``strong second law,'' that express precise (but inequivalent) senses of increase of $S_\qB$. 

\begin{rmk}[Status and relevance of the two conditions]\label{rem:super}
The question arises, what do we even mean by saying that $S_\qB$ increases with $t$, given that $\psi_t$ can be a superposition of different values of $S_\qB$? At this question, the foundations of statistical mechanics touch the foundations of quantum mechanics, where questions have been debated \cite{Tum_22} such as, what is reality like when the wave function is a superposition of macroscopically different contributions (like, e.g., Schr\"odinger's cat)? Different theories of quantum mechanics (such as Bohmian mechanics \cite{DT}, spontaneous collapse theories \cite{GB25}, or many-world theories \cite{Wal12,AGTZ11}) lead to different conclusions about this reality and, as a consequence, about actual entropy values \cite[Sec.~3.1]{GLTZ_19}. A history $t\mapsto S_\qB(t)$ of actual values of $S_\qB$ may arise, either from the hidden variables, or from spontaneous collapses breaking up the superposition, or in every branch of the wave function; of that curve $t\mapsto S_\qB(t)$, we can then ask whether it is (predominantly) increasing during the time interval $[0,T]$, where $T$ is the time scale of reaching thermal equilibrium.
The two conditions defined in this section can be applied in each of these theories of quantum mechanics, as they do not talk about the hidden variables or collapses. And yet, it is plausible that both are necessary conditions for the increase of $S_\qB(t)$ with $t$ in each of these theories, and that the strong second law is also sufficient.
$\hfill \diamond$
\end{rmk}

\subsection{Weak Second Law}\label{sec:WSL}

We now introduce the concept of \emph{right-moving distributions}. Let
\begin{equation}
    \SSS(\Hilbert)=\{ \psi \in \Hilbert : ||\psi || =1 \}
\end{equation}
denote the unit sphere of $\Hilbert$. For every $\psi_0\in\SSS(\Hilbert)$ and every time $t$, $\psi_t$ defines a distribution of the macroscopic superposition weights
\begin{equation}\label{eq:weights}
    p_\nu(t):=||P_\nu \psi_t ||^2
\end{equation}
over $\mathscr{N}$. The collection $(p_\nu(t))_{\nu \in \mathscr{N}}$ can be considered as the macroscopic description of $\mathscr{S}$ at time $t$. Next, we define an order on the set $\mathscr{N}$ of macro states by 
\be
\nu<\nu' :\Leftrightarrow d_{\nu}<d_{\nu'} \:.
\ee
(For simplicity, we neglect the case in which $d_\nu=d_{\nu'}$ for some $\nu\neq \nu'$.)

\begin{figure*}[t] % Das Sternchen * sorgt fŸr die doppelte Spaltenbreite
    \centering
    {\includegraphics[width=0.24\textwidth]{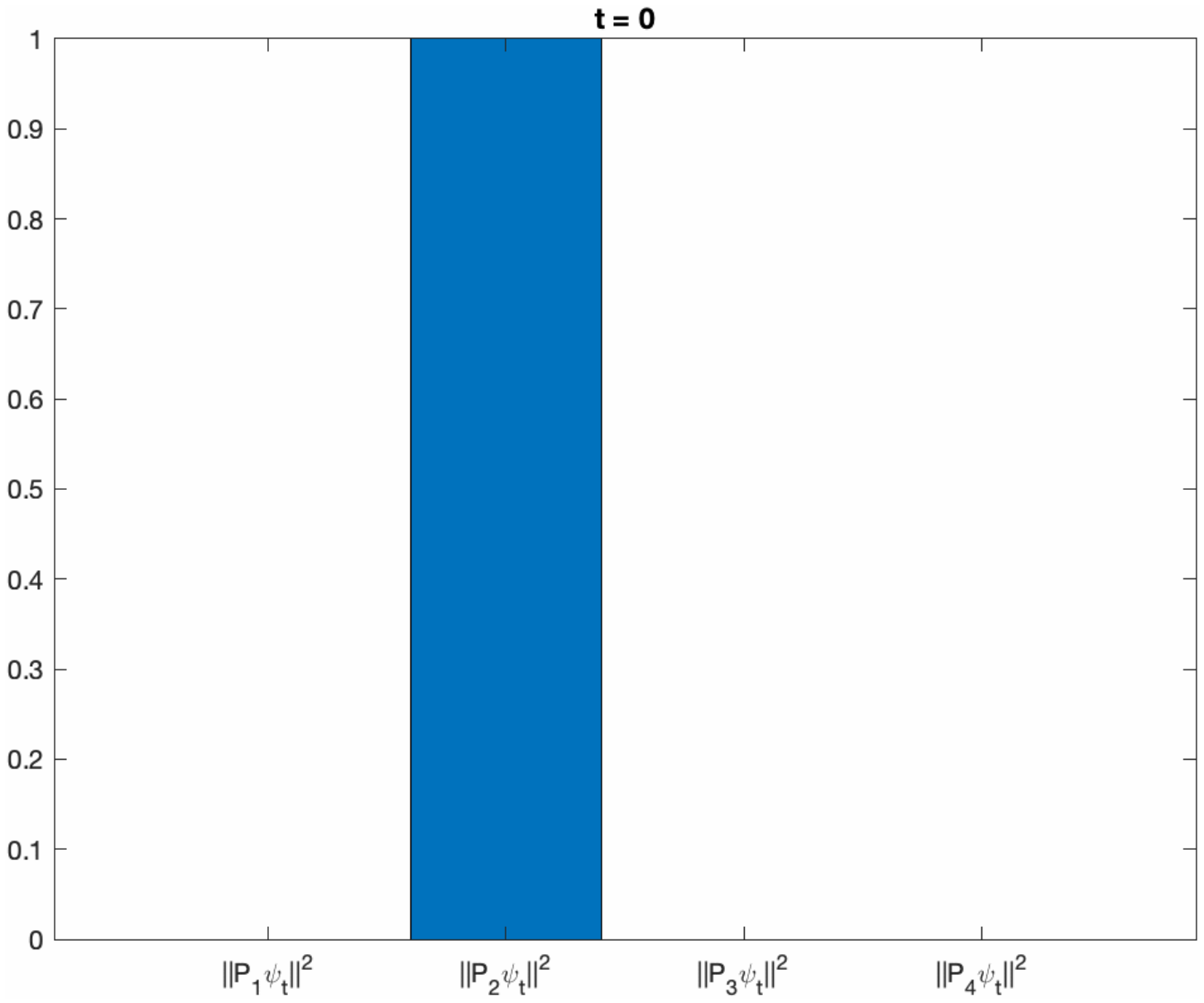}}
    %\subfloat[Zweite Grafik\label{fig}]
    {\includegraphics[width=0.24\textwidth]{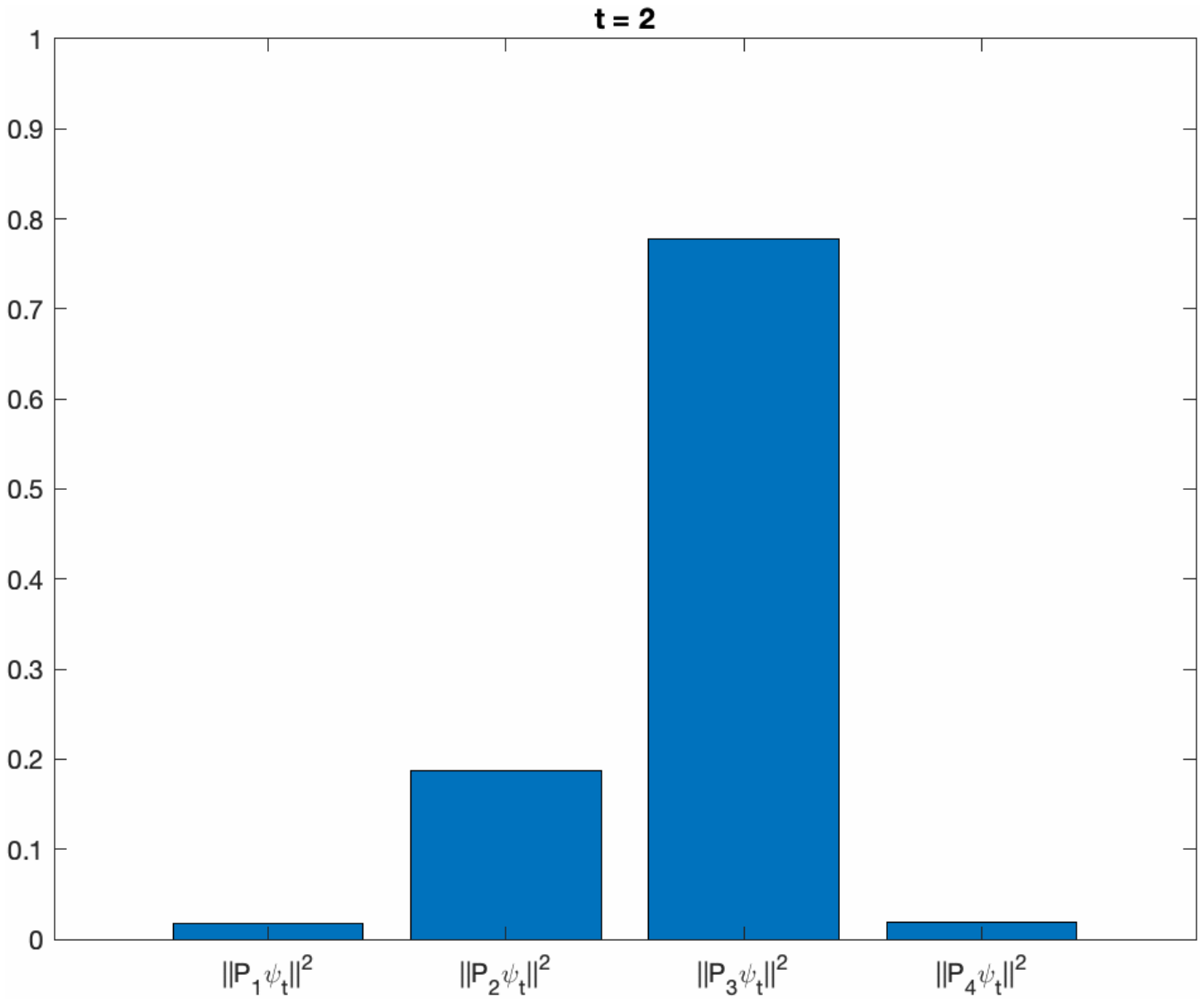}}
    %\subfloat[Zweite Grafik\label{fig}]
    {\includegraphics[width=0.24\textwidth]{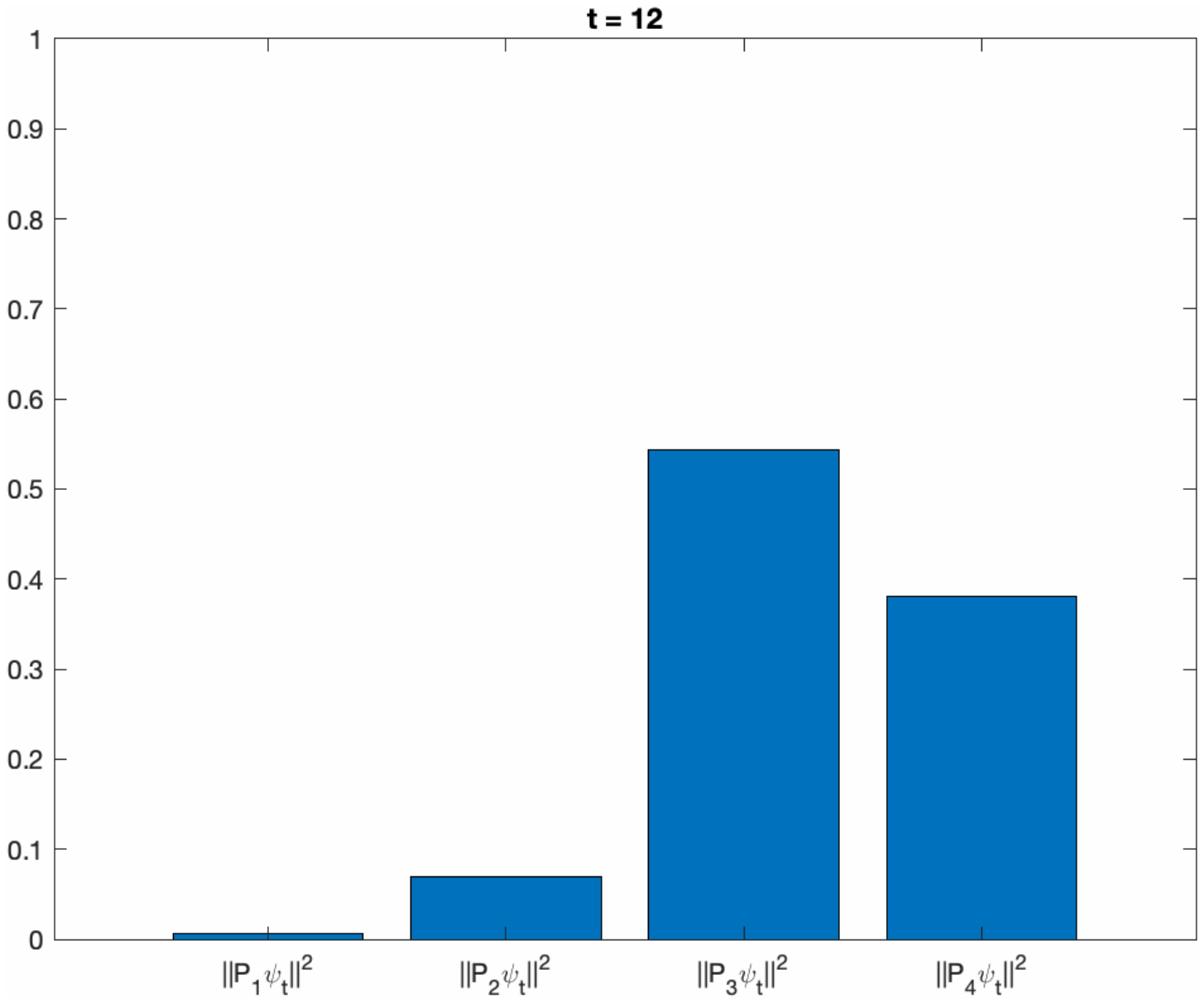}}
    %\subfloat[Zweite Grafik\label{fig}]
    {\includegraphics[width=0.24\textwidth]{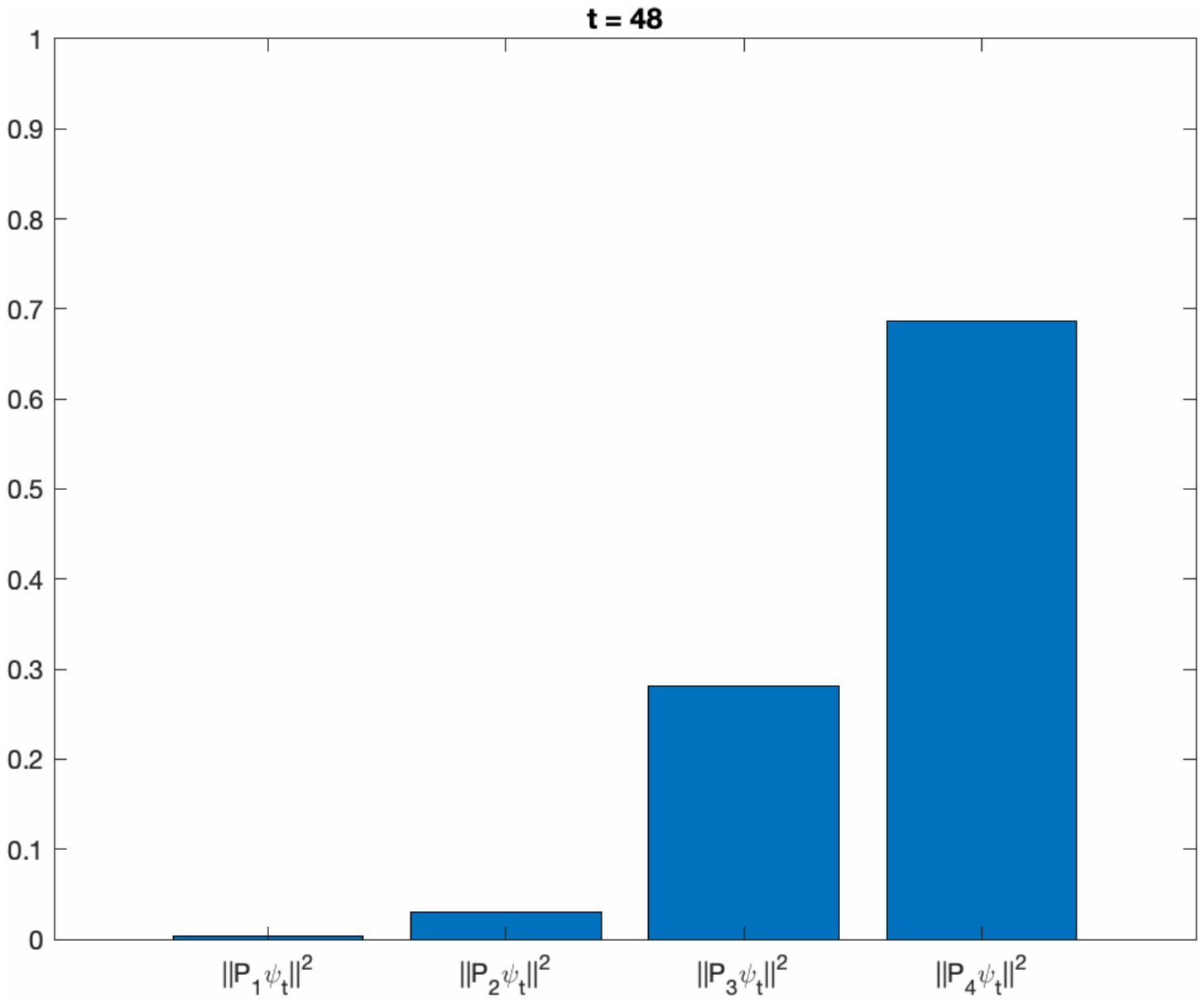}}
    \caption{\footnotesize Histograms of $p_\nu(t)=\|P_\nu\psi_t\|^2$ at $t=0,2,12,48$ in arbitrary units, obtained through numerical simulation (courtesy of Cornelia Vogel) for a decomposition $\Hilbert=\Hilbert_\mc$ into macro spaces $\Hilbert_1\oplus \Hilbert_2 \oplus \Hilbert_3 \oplus \Hilbert_4$ with dimensions $d_1=2,d_2=20,d_3=200,d_4=2000$, a typical $\psi_0\in\SSS(\Hilbert_2)$, and a Hamiltonian that is a random band matrix. One can see that the family of histograms is approximately right-moving. The same data were displayed in \cite[Fig. 1]{TTV_23} in a different way.
    \label{fig:RMDEx}}
\end{figure*}

Now for each $t\geq 0$, $(p_\nu(t))_{\nu \in \mathscr{N}}$ can be visualized as a histogram on $\mathscr{N}$ with a column for each macro state $\nu$. As time evolves, the histogram changes and probability is transported to different macro states $\nu$; correspondingly for each $\nu$, the column $p_\nu(t)$ may shrink or rise with $t$. Thus, for $\nu$ (and $S_\qB(\nu)$) to increase with time, probability has to be moved ``to the right'' (see Figure~\ref{fig:RMDEx}).
    
Let us make this precise by means of two definitions. First, we say that the probability distribution $\tilde{p}$ \emph{lies to the right of} $p$ and write $p\preceq \tilde{p}$ if and only if
\begin{equation}
     \sum_{\substack{\nu\in\mathscr{N}\\\nu\geq\nu'}}p_{\nu}\leq \sum_{\substack{\nu\in\mathscr{N}\\\nu\geq\nu'}}\tilde{p}_{\nu} \quad \forall \nu'\in \mathscr{N}.
\end{equation}
(This concept is also known as the ``usual stochastic order'' \footnote{\url{http://en.wikipedia.org/wiki/Stochastic_ordering}} of random variables $X,\tilde{X}$ with distributions $p,\tilde{p}$.) Equivalently, $p\preceq \tilde{p}$ if and only if $\tilde{p}$ can be obtained from $p$ by only moving amounts of probability to the right or, in other words, if and only if there is a pair $(X,\tilde{X})$ of random variables, each with values in $\mathscr{N}$, such that the marginal distribution of $X$ is $p$, the marginal distribution of $\tilde{X}$ is $\tilde{p}$, and always $X\leq \tilde{X}$.

Second, a $t$-indexed family of probability distributions $(p(t))_{t\in[0,T]}$ on $\mathscr{N}$ is \emph{right-moving} if and only if
\begin{equation}\label{eq:RMDcond}
    \forall t_{1}\leq t_{2}: p(t_{1})\preceq p(t_{2}),
\end{equation}
where $t_{1},t_{2}\in[0,T]$. Equivalently, $(p(t))_{t\in[0,T]}$ is right-moving if and only if there is a stochastic process $(X_t)_{t\in[0,T]}$ such that $X_t$ has distribution $p(t)$ and $X_{t_1}\leq X_{t_2}$ whenever $t_1\leq t_2$. Another equivalent condition, given that $p_\nu(t)$ is a differentiable function of $t$, is that
\be\label{eq:RMDcond2}
\frac{\mathrm{d}}{\mathrm{d}t}\sum_{\substack{\nu\in\mathscr{N}\\\nu\geq\nu'}}p_{\nu}(t) \geq 0
\ee
for all $\nu'\in\mathscr{N}$ and $t\in[0,T]$.

As is well known from classical statistical mechanics, violations of the second law are not impossible. Also here, several types of violations must be expected to occur: first, due to Poincar\'e recurrence, but this happens only after exorbitantly long waiting times of the order $e^D$ \cite{CRV_22} \footnote{More precisely, 
%Certainly due to Poincar\'e  recurrence entropy will not stay at its maximal value for all times, but for long enough times. Recurrence times, at least classically grow with the number of particles. For closed quantum systems similar statements are expected to hold \cite{CRV_22}. Further 
in closed quantum systems the recurrence time 
%$t_\text{rec}$ scales as $t_\text{rec}\approx \mathcal{O}(\exp(d_\text{eff}))$, 
is of order $\exp(d_\text{eff})$, where $d_\text{eff}=1/ \sum_n |\langle \psi_0 | \phi_n \rangle|^4$ is the effective number of energy eigenstates $\phi_n$ contributing to $\psi_0$ and is typically comparable to $D$.};
second, tiny amounts of probability will be transported to smaller subspaces; third, very infrequently (i.e., after long waiting times), also not-so-tiny amounts of probability may be transported to smaller subspaces, but this is expected to happen in an uncontrollable and practically unpredictable way and only, as stated, after long waiting times; finally, a tiny set of exceptional wave functions $\psi_0\in\SSS(\Hilbert_{\nu_0})$ may feature large violations of the second law, which is why we consider typical $\psi_0\in\SSS(\Hilbert_{\nu_0})$. As a consequence, one can only expect that \eqref{eq:RMDcond} holds \emph{approximately}; we call such distributions  \emph{approximately right-moving distributions} and give more detail below in Section~\ref{sec:ApproxInc}.

We are now able to state a version of the second law:

\bigskip

\noindent \textbf{Weak Second Law (WSL):}\label{quote:WSL}
{\it We say that a system with Hamiltonian $H$, macro spaces $\Hilbert_\nu$, initial macro state $\nu_0$, and $\psi_0\in \SSS(\Hilbert_{\nu_0})$ satisfies the weak second law if the family $p_{\nu}(t)=\|P_\nu\psi_t\|^2$ is an approximately right-moving distribution up to time $T$, where $T$ is the time scale of reaching MATE.}

\subsection{Strong Second Law}\label{sec:SSL}

There is another, even stronger, way to understand the increase of $S_\qB$, namely via the probability current between two macro spaces $\Hilbert_\nu$ and $\Hilbert_{\nu'}$ for $\nu\neq\nu'$,
\begin{equation}
\begin{aligned}\label{eq:Current}
    \mathcal{J}_{\nu\nu'}^{\psi_{0}}(t)&:= 2 \, \mathrm{Im} \langle \psi_{t} \vert P_{\nu} H P_{\nu'} \vert \psi_{t} \rangle, \\
\end{aligned}
\end{equation}
where $\psi_0 \in \mathbb
{S}(\mathscr{H}_{\nu_{0}})$ (with an initial macro state $\nu_0$ possibly different from $\nu$ and $\nu'$). This quantity represents the amount of probability transported per time from $\Hilbert_{\nu'}$ to $\Hilbert_\nu$ minus the other way around.
Note that $\mathcal{J}_{\nu'\nu}=-\mathcal{J}_{\nu\nu'}$, in particular $\mathcal{J}_{\nu\nu}=0$.

The current \eqref{eq:Current} is related to \eqref{eq:weights} via a discrete continuity equation 
\begin{align}\label{ContEq}
 \frac{\mathrm{d}}{\mathrm{d}t} p_{\nu}(t) = \sum_{\nu'\in\mathscr{N}} \mathcal{J}_{\nu\nu'}^{\psi_0}(t).
\end{align}
The increase of $S_\qB$ can now be expressed via a directedness of the current from macro spaces with small dimension toward  macro spaces with larger dimension: 

\bigskip

\noindent    \textbf{Strong Second Law (SSL):}\label{quote:SSL} 
{\it We say that a system with Hamiltonian $H$, macro spaces $\Hilbert_\nu$, initial macro state $\nu_0$, and $\psi_0\in \SSS(\Hilbert_{\nu_0})$
satisfies the strong second law if the discrete current \eqref{eq:Current} points overwhelmingly from smaller $\mathscr{H}_{\nu'}$ to larger $\mathscr{H}_{\nu}$:
\begin{equation}\label{eq:CurPoi}
    \mathcal{J}_{\nu\nu'}^{\psi_0} (t) \geq 0 \quad \text{for} \, \, \nu' < \nu
\end{equation}
for $t\in[0,T]$, where $T$ is the time scale of reaching MATE.} 

\bigskip

The SSL implies the WSL, but for $\# \mathscr{N}>2$ not vice versa (see Appendix~\ref{app:SSL2WSL} for the proof). In fact, the WSL is equivalent to the existence of \emph{some} quantities $\mathcal{J}_{\nu\nu'}(t)=-\mathcal{J}_{\nu'\nu}(t)$, not necessarily given by \eqref{eq:Current}, satisfying \eqref{ContEq} and \eqref{eq:CurPoi} (see Appendix~\ref{app:WSLcurrent}).

\begin{rmk}[Markovian approximation] 
    A related statement     was proposed as a conjecture by van Kampen \cite{vKam_54} and recently argued for by Strasberg et al. \cite{SWGW_23}. In the version relevant here, it says that $p_\nu(t)=\|P_\nu\psi_t\|^2$ agrees with the distribution at time $t$ of a Markov jump process on $\mathscr{N}$ with ($t$-independent) jump rates
    \begin{equation}\label{eq:locdetajumprates}
	\sigma_{\nu\nu'} = \frac{1}{\tau d_{\nu'}} \mathrm{tr}(e^{i H \tau} P_\nu e^{-i H \tau} P_{\nu'})
    \end{equation}
    for $\tau>0$ small (such that the expression does not depend much on $\tau$), but not too small (or else $\sigma_{\nu\nu'}\approx 0$). (In fact, \eqref{eq:locdetajumprates} equals $\tau^{-1}$ times the probability that a quantum measurement of $P_\nu$ at time $\tau$ yields outcome 1 on a system starting in a typical $\psi_0\in\SSS(\Hilbert_{\nu'})$.) As a consequence of this conjecture, $p_\nu(t)$ obeys the master equation
    \begin{equation}\label{eq:locdetweights}
	   \frac{\mathrm{d}p_\nu(t)}{\mathrm{d}t} = \sum_{\nu'} (\sigma_{\nu\nu'}p_{\nu'}(t) - \sigma_{\nu'\nu}p_\nu(t)) \,.
    \end{equation}
    Both van Kampen and Strasberg et al.\ further argue for an extension of this conjecture, called the \emph{hypothesis of local detailed balance}, which says that
    \begin{equation}\label{eq:hlocdetbal}
	   \frac{\sigma_{\nu\nu'}}{\sigma_{\nu'\nu}} \approx \frac{d_\nu}{d_{\nu'}}.
    \end{equation}
    This (more or less) implies the WSL because for macro states $\nu'<\nu$ with significantly different entropy values, $d_{\nu'}\ll d_\nu$, so by \eqref{eq:hlocdetbal} $\sigma_{\nu'\nu}\ll \sigma_{\nu\nu'}$, which means that probability gets transported predominantly from smaller to larger macro states (unless already $p_{\nu'}(t)\ll p_\nu(t)$ in the same proportion as $d_{\nu'}\ll d_\nu$). Eq.~\eqref{eq:hlocdetbal} can be confirmed for certain random matrix models (see Remark~\ref{rmk:locdetWig}), but it remains open whether it is valid more generally.$\hfill \diamond$
\end{rmk}

\subsection{Approximate Entropy Increase}\label{sec:ApproxInc}

We have formulated versions of the second law that permit minor violations of entropy increase. Now we make precise what we mean by ``approximately'' right-moving in the WSL (Section~\ref{quote:WSL}) and ``overwhelmingly pointing to larger $\Hilbert_\nu$'' in the SSL (Section~\ref{quote:SSL}). In Section~\ref{sec:2Wig}, we illustrate this concept using a specific random matrix model. Let $x^+=\max(x,0)$ denote the positive part of $x\in\RRR$, let
\begin{equation}\label{eq:defndeltaell}
	\delta_{\ell}(t):= \int_{0}^{t} \mathrm{d}t' \, \sum_{\substack{\nu,\nu'\in\mathscr{N}\\\nu < \nu'}} \mathcal{J}_{\nu\nu'}^{+}(t') 
\end{equation}
be the total amount of probability transported before $t$ to the left  (i.e., to smaller subspaces, violating the second law), and let
\begin{equation}\label{eq:defndeltar}
	\delta_{r}(t):= \int_{0}^{t} \mathrm{d}t' \, \sum_{\substack{\nu,\nu'\in\mathscr{N}\\\nu > \nu'}} \mathcal{J}_{\nu\nu'}^+(t') 
\end{equation}
be the total amount of probability transported before $t$ to the right (i.e., to larger subspaces). Equivalently, $\delta_\ell(t)$ and $\delta_r(t)$ are the expected number of jumps to the left (right) during $[0,t]$ of a stochastic process $(X_t)$ in $\mathscr{N}$ having distribution $(p_\nu(t))_\nu$ and jump rates $\sigma_{\nu\nu'}(t)=\mathcal{J}^+_{\nu\nu'}(t)/p_{\nu'}(t)$ for every $t$. 

We say that the SSL holds \emph{approximately} whenever
\begin{equation}\label{eq: deltap}
	\delta_p(t) := \frac{\delta_\ell(t)}{\delta_r(t)} \ll 1
\end{equation}
for most $t\in[0,T]$, where $T$ is again the time scale of reaching MATE. (An alternative definition, inequivalent but similar in spirit, could be based on the expected amounts of decrease/increase of entropy, see Appendix~\ref{app:defdeltaS}.)

Similarly, we say that the WSL holds \emph{approximately} whenever there exist quantities $\mathcal{J}_{\nu\nu'}(t)=-\mathcal{J}_{\nu'\nu}(t)$, not necessarily given by \eqref{eq:Current} but satisfying \eqref{ContEq}, for which \eqref{eq: deltap} holds for most $t\in[0,T]$.

\section{Example 1: Low-Dimensional Subspace in Random Direction}
\label{sec:ex1}\label{sec:simpleexample}

Let $\mathscr{H}_{\mathrm{mc}}=\mathscr{H}_{1}\oplus\mathscr{H}_{2}$ with two macro spaces $\Hilbert_1,\Hilbert_2$. Let us begin with the extreme case in which $d_{1}=1$ and $d_2 =\dim \Hilbert_\mc-1= D -1$. Let $H=\sum_{n}E_{n}\vert \phi_n \rangle \langle \phi_n \vert$ be the Hamiltonian. We start with $\psi_{0}\in\mathbb{S}(\mathscr{H}_{1})$ and ask ``how much'' of $\psi_t$ will be left in $\mathscr{H}_1$ after some time $t>0$. The random direction of $\Hilbert_1$ can be implemented by choosing $\psi_0$ uniformly distributed on $\SSS(\Hilbert_\mc)$; since $D\gg 1$, the coefficients $c_{n}$ in the energy expansion $\psi_{0}=\sum_{n}c_{n}\vert \phi_n \rangle$ are approximately independent complex Gaussian random variables with mean 0 and variance $1/D$.
Thus, as the energy levels are very close,
\begin{align}\label{1dOverlap}
    \langle \psi_{0} \vert \psi_{t}\rangle 
    &= \sum_n c_n^* e^{-itE_n} c_n\\
    & \approx \frac{1}{D} \sum_{n}e^{-itE_{n}}\\
    &\approx \int_{\mathbb{R}} \mathrm{d}E \, \rho(E) \, e^{-itE} \label{contapprox}\\
    &=\sqrt{2\pi} \: \hat{\rho}(t)
\end{align}
which is, up to a factor, the Fourier transform $\hat{\rho}(t)$ of the density of states $\rho(E)$ (similar approximations were made already in \cite{Rei_19,Rei_16}). That is, with $P_{1}$ the projector on $\mathscr{H}_{1}$, 
\begin{equation}\label{P1hatrho}
        \norm{P_{1}\psi_{t}}^2=\vert \langle \psi_{0} \vert \psi_{t} \rangle \vert^2 \approx 2\pi\, \vert \hat{\rho}(t) \vert^2.
    \end{equation}
The Riemann-Lebesgue lemma asserts that for a function $f$ with $\int_\RRR |f|<\infty$ (such as $\rho$), $\hat{f}(t)$ tends to 0 as $t\to \pm \infty$. This implies that for $t$ that is large (so we can replace $\hat\rho(t)$ by $\lim_{t\to\infty} \hat\rho(t)$) but not too large (so the continuum approximation \eqref{contapprox} is still valid), $\psi_{t}$ will overwhelmingly lie in the bigger subspace $\Hilbert_2$. (The Poincar\'e recurrence theorem tells us that there exist larger $t$ for which $\psi_t$ will be far from $\Hilbert_2$ and close to $\Hilbert_1$ again; the recurrence time tends to $\infty$ in the limit $D\to\infty$ that corresponds to the continuum approximation \eqref{contapprox}.)

Up to now we have left open what $\rho(E)$ looks like. As a simple example (actually proven in Theorem~\ref{thm:1} below), if we take $H$ to be a GUE random matrix in $\Hilbert_\mc$, then since $D\gg 1$, $\rho(E)$ is with very high probability approximately given by the Wigner semicircle density
\begin{equation}\label{eq:WigSemi}
\rho_{\mathrm{sc}}(E) = \frac{1}{2\pi} \sqrt{4-E^2} \, \mathds{1}_{[-2,2]}(E),
\end{equation}
which leads to
\begin{equation}
    \norm{P_{1}\psi_{t}}^2 \approx \frac{J_{1}(2t)^2}{t^{2}}
\end{equation}
with $J_1(\cdot)$ the first Bessel function of the first kind.
This function is not strictly monotonic, and thus the distribution $p_\nu(t)$ not strictly right-moving; but as shown in Appendix~\ref{app:PfCor1} for the purpose of deriving Statement~\ref{cor:1}, it is approximately right-moving within a tolerance of $\delta_p(T)\leq 2.5\%$.

We believe, however, that in realistic examples, the tolerance $\delta_p(T)$ in the WSL will be way smaller. A realistic density of states, cut off at some energy value $E_{\max}$ to resemble a micro-canonical energy shell, would satisfy the hypotheses of the following theorem.

\begin{thm}\label{thm:Fourier}
Suppose the function $\rho(E)\geq 0$ vanishes outside $[0,E_{\max}]$, satisfies 
\be
\rho(0)=\rho'(0)=0
\ee
(where $\rho'=\mathrm{d}\rho/\mathrm{d}E$), is three times continuously differentiable in $[0,E_{\max}]$, and satisfies
\be
\rho'\ge0,\quad \rho''\ge0,\quad \rho'''\ge0
\quad\text{on }(0,E_{\max})\,.
\ee
Then
\be\label{rhodecrease}
\frac{\mathrm{d}}{\mathrm{d}t}|\hat\rho(t)|^2 \leq 0 \text{ for }t>0 \,.
\ee
\end{thm}

We give the proof in Appendix~\ref{app:Fourier}. The hypotheses are satisfied in particular for the $\rho$ of an ideal quantum gas in a box, cut off at $E_{\max}$,
\be\label{powerdensity}
\rho(E) = C\, E^{\alpha} \, \mathds{1}_{[0,E_{\max}]}(E)
\ee
with normalizing constant $C>0$, provided the constant exponent $\alpha$ is $\geq 3$. (For a gas in a box, $\alpha=3N/2-1$ with $N$ the particle number, so $\alpha\gg 1$.)

Eq.~\eqref{rhodecrease} entails that violations of being right-moving cannot arise from oscillations in the function $|\hat\rho(t)|^2$ but only from the error in \eqref{P1hatrho}, which tends to 0 as $D\to\infty$, so the WSL holds with very small tolerance.

Let us now turn to the case in which $d_1>1$, but still $d_1/D\ll 1$. A similar consideration as before (see Appendix~\ref{app:d1}) shows the following.

\begin{sta}\label{sta:d1}
For fixed $d_1$, $\Hilbert_1$ a purely random subspace of dimension $d_1$, purely random $\psi_0\in\SSS(\Hilbert_1)$, any $t$ and sufficiently large $D$ (or conversely, when a large $D$ is given, for any $t$ small compared to the recurrence time $\mathcal{O}(e^D)$),
\be\label{ex1d1}
\|P_1 \psi_t \|^2 \approx 2\pi \, |\hat\rho(t)|^2
\ee
with very high probability.
\end{sta}

\section{Example 2: GUE and Wigner Matrices}
\label{sec:ex2}\label{sec:2Wig}

The result in this section is based on work of Cipolloni et al.~\cite{CES_22}; it shows that the WSL and SSL hold for any choice of the $\Hilbert_\nu$ when $H$ is a random GUE matrix, or more generally a Wigner matrix. A \emph{Wigner matrix} is (for us) a complex $D\times D$ Hermitian random matrix $H=A+A^\dagger$, $A=(A_{ij})$, where the $\mathrm{Re}\, A_{ij}$ and $\mathrm{Im}\, A_{ij}$ are all independent and identically distributed with mean 0, variance $1/4D$ (so that all entries $H_{ij}$ of $H$ have $\EEE H_{ij}=0$ and $\EEE|H_{ij}|^2=1/D$), and all higher moments finite \footnote{The definition of Wigner matrix used in \cite{CES_22}, and thus what is needed to ensure their Theorem 2.1 and our Theorem~\ref{thm:1}, is somewhat broader: of $H_{ij}=H_{ji}^*$ with $i<j$ it is required that they are independent and identically distributed in $\mathbb{C}$ with $\EEE H_{ij}=0$, $\EEE|H_{ij}|^2=1/D$, $\EEE H_{ij}^2\in\mathbb{R}$ (which is automatic when $\mathrm{Re}\,H_{ij}$ and $\mathrm{Im}\, H_{ij}$ are independent with mean 0), and finite higher moments; furthermore, all $H_{ii}$ are independent and identically distributed in $\RRR$ (but the distribution need not be related to that of the off-diagonal entries) with $\EEE H_{ii}=0$, $\EEE H_{ii}^2= (\mathrm{const.})/D$, and finite higher moments.}.
We write $\PWig$ for any Wigner distribution.
Wigner matrices have several universal properties \cite{EY_16,Erd_19,CES_21b}. For instance, it is well known \cite{Erd_19,EY_16} that for $D\gg 1$ the empirical distribution of the eigenvalues of $H$ is with very high probability given by the semicircle density \eqref{eq:WigSemi}.

Let us make a conceptual remark here. 

\begin{rmk}[Typicality]
    The appeal to typicality \cite{Gol_12,LR_15,Wil_22,Laz_23} implicit in modeling $H$ as a random matrix has several reasons: (i) the ``real'' $H$ is complicated, and a typical (or generic) $H$ is easier to study \cite{vNeu_29,Wig_55,GLMTZ_10n,Rei_16,DR_20,RSTTV_25}, (ii) entropy increase should be universal and  not depend on the precise details of $H$, (iii) a typicality result is informative, and it is reasonable to assume that it applies to a specific $\mathscr{S}$ unless good reasons indicate the opposite \cite{BGVZ_25,GLMTZ_10n}, (iv) for related thermal behavior such as thermalization, typicality reasoning has yielded relevant insights \cite{GLMTZ_10a,Rei_15,Rei_16,GE_16,Mori_18,DR_20,CES_21,CES_22,ER_24,RSTTV_25}.
Typicality statements are phrased in the language of probability theory, but it is important to note that there is not necessarily any ``intrinsic'' randomness. Rather, a typicality statement about $H$ means that the set of ``bad'' Hamiltonians is small (with respect to a suitable measure).$\hfill\diamond$
\end{rmk}

\subsection{WSL and SSL for Wigner Matrices}\label{sec:}

Before we compute the weights \eqref{eq:weights} and the current \eqref{eq:Current}, let us introduce some notation. The goal is to make a typicality statement about $H$.
Let $\text{Herm}(\Hilbert)$ denote the set of Hermitian operators on $\Hilbert$ and $\varepsilon>0$. We say a statement $s(H)$ about $H$ is true \emph{for $(1-\varepsilon)$-most} $H$ if and only if 
\begin{equation}\label{eq:WigTyp}
	\PWig\left(  \{H \in \text{Herm}(\Hilbert) : s(H) \, \text{holds} \} \right) \geq 1- \varepsilon.
\end{equation}
With this we can formulate the following:

\begin{thm}\label{thm:1}
    For any $\varepsilon,\delta>0$, there exists $D_0=D_0(\varepsilon,\delta)>0$ such that for all $D>D_0$, the following holds.
    Let $\Hilbert_\mc$ be a Hilbert space of dimension $D$, $\bigoplus_\nu \Hilbert_\nu$ any orthogonal decomposition of it, $\nu,\nu',\nu_0$ arbitrary indices, and $H$ a random operator in $\Hilbert_\mc$ given, in some orthonormal basis aligned with the $\Hilbert_\nu$, by a GUE or Wigner matrix. For any $\psi_0 \in \SSS(\Hilbert_{\nu_0})$ and any $0<t\ll D^{1/7}$ \footnote{The unit of time here is $\hbar$ divided by the energy unit of the semicircle density \eqref{eq:WigSemi}.}, $(1-\varepsilon)$-most $H$ are such that
    \begin{align}\label{WeightsWignerLargeD}
		\W&\approx \frac{d_{\nu}}{D} + \frac{J_1(2t)^2}{t^2} \left( \delta_{\nu\nu_{0}}-\frac{d_{\nu}}{D}\right),\\
	\label{eq:WignerCurrentSSL}
        \mathcal{J}_{\nu\nu'}^{\psi_{0}}(t) 
        &\approx\frac{4}{D}(d_{\nu}\delta_{\nu_{0}\nu'}-d_{\nu'}\delta_{\nu_{0}\nu}) \frac{J_1(2t)J_2(2t)}{t^2},
    \end{align}
where ``$\approx$'' allows an error of size $\delta$ and $J_k(\cdot)$ is the $k$-th Bessel function of the first kind.
\end{thm}

We give the proof in Appendix~\ref{sec:DerWSLWig}. From Theorem~\ref{thm:1}, we obtain (though not in a mathematically rigorous way) the WSL and the SSL for Wigner matrices: 

\begin{sta}\label{cor:1}
Consider any system satisfying the assumptions of Theorem~\ref{thm:1} for sufficiently small $\varepsilon,\delta$ (and thus sufficiently large $D$), and suppose $\nu_0<\eq$ with \eqref{deqD1}. Then the WSL and SSL hold, up to the time $T$ of reaching MATE, within a tolerance (cumulative error) of $\delta_p(T)\leq 2.5\%$.
\end{sta}

We derive this in Appendix~\ref{app:PfCor1}. 

\begin{rmk}[Local detailed balance for Wigner matrices]\label{rmk:locdetWig}
  Let $\nu,\nu'$ be two macro states and let $H$ be a Wigner matrix. Then for sufficiently large $\tau$, 
  \begin{equation}\label{eq:locdetWig}
		\frac{\sigma_{\nu\nu'}}{\sigma_{\nu'\nu}} \approx \frac{d_\nu}{d_{\nu'}}
	\end{equation}
in line with \eqref{eq:hlocdetbal}. We derive this in Appendix~\ref{app:locdetWig}.\hfill$\diamond$	
\end{rmk}

\section{Example 3: Random Block-Band Matrix}
\label{sec:ex3}

A random block-band matrix is still a toy model, but more realistic. 
For this model, let $\Hilbert=\Hilbert_\mc=\Hilbert_1 \oplus \Hilbert_2 \oplus \Hilbert_3$ with $d_1\ll d_2 \ll d_3=d_\eq$, $d_1+d_2+d_3=D$.
Take $H$ to be a \emph{Wigner-type} Hermitian random matrix, i.e., $H=A+A^\dagger$, $A=(a_{ij})$, where the $a_{ij}$ are independent, $\EEE a_{ij}=0$, and $\EEE|a_{ij}|^2=v_{ij}$ with
\be
V= (v_{ij})=\begin{pmatrix} V_{11} & V_{12} & 0 \\
V_{21}&V_{22} & V_{23}\\
0&V_{32} & V_{33}\end{pmatrix}\,,
\ee
where each block $V_{\nu\nu'}$ has all entries equal. The results of Erd\H os, Henheik, and Vogel \cite{EHV_25} imply the following:

\begin{sta}
For $d_2/d_1= d_3/d_2$, a suitable choice of the entries of $V_{\nu\nu'}$ (that allows explicit calculations), and any $\psi_0\in\SSS(\Hilbert_{\nu_0})$, it has probability near 1 that for times $t$ that are large ($> \mathcal{O}(d_2/d_1)$) but not too large ($\ll$ recurrence times $\mathcal{O}(e^D)$), the $\|P_\nu \psi_t\|^2$ evolve monotonically (up to errors that vanish as $D\to\infty$) towards $d_\nu/D$. In particular, for $t\gg d_2/d_1 \gg 1$, the WSL holds (right-moving distribution), but thermalization is not ultrafast.
\end{sta}

\section{Comparison and Outlook}
\label{sec:outlook}

Various statements are considered as formulations of the second law of thermodynamics in the literature. Some focus on thermal equilibrium states and transitions between them \cite{Mei_24}, some on the impossibility of certain (perpetual motion) machines \cite{GHT_13,ST_25,HU25,Chiba26}. We expect that such statements can ultimately also be derived from the increase of Boltzmann entropy.

For our approach to the second law, we see the main challenge for the future in providing a derivation of the second law from quantum mechanics of the following kind: to formulate a condition on $H$ and the $\Hilbert_\nu$, to provide a mathematical proof that this condition implies the weak or strong second law as stated in Section~\ref{sec:2BQF} above, and to argue that realistic macroscopic systems satisfy this condition.

\appendix

\phantomsection
\section*{Appendix}\label{sec:appendix}

\section{Strong Second Law Implies Weak Second Law}
\label{app:SSL2WSL}

From \eqref{eq:CurPoi} we obtain for any $\nu''$ that (omitting the $t$-dependence for ease of notation)
\begin{align}
&\frac{\mathrm{d}}{\mathrm{d}t} \sum_{\substack{\nu\in\mathscr{N}\\\nu\geq\nu''}}p_{\nu}
\stackrel{\eqref{ContEq}}{=} \sum_{\substack{\nu\in\mathscr{N}\\\nu\geq\nu''}}\sum_{\nu'\in\mathscr{N}} \mathcal{J}_{\nu\nu'}\\
&= \sum_{\substack{\nu\in\mathscr{N}\\\nu\geq\nu''}}\Biggl( \sum_{\substack{\nu'\in\mathscr{N}\\\nu'<\nu''}}
+ \sum_{\substack{\nu'\in\mathscr{N}\\\nu''\leq \nu'\leq \nu}}
+ \sum_{\substack{\nu'\in\mathscr{N}\\\nu'>\nu}} \Biggr) \mathcal{J}_{\nu\nu'}\\
&= \sum_{\substack{\nu,\nu'\in\mathscr{N}\\\nu'<\nu''\leq\nu}} \underbrace{\mathcal{J}_{\nu\nu'}}_{\geq 0}
+ \Biggl(\sum_{\substack{\nu,\nu'\in\mathscr{N}\\\nu''\leq \nu'\leq \nu}}
+ \sum_{\substack{\nu,\nu'\in\mathscr{N}\\\nu''\leq \nu<\nu'}} \Biggr) \mathcal{J}_{\nu\nu'}\\
\intertext{[now rename $\nu \leftrightarrow \nu'$ in the last sum]}
&\geq \sum_{\substack{\nu,\nu'\in\mathscr{N}\\\nu''\leq \nu'< \nu}}
\Bigl( \underbrace{\mathcal{J}_{\nu\nu'}+\mathcal{J}_{\nu'\nu}}_{=0} \Bigr)=0
\end{align}
so, by \eqref{eq:RMDcond2}, $p$ is right-moving. 
Conversely, given values of $p_\nu(t)$ for all $\nu$
and $t$ do not fix the $\mathcal{J}_{\nu\nu'}(t)$; 
they fix the row sums by \eqref{ContEq} but leave the 
freedom of adding any anti-symmetric matrix $\mathcal{A}_{\nu\nu'}(t)$ 
with zero row sums, which can be non-zero and can change 
the sign of $\mathcal{J}_{\nu\nu'}(t)$.

\section{WSL and Currents}
\label{app:WSLcurrent}

We prove the statement of Section~\ref{sec:SSL} that the WSL is equivalent to the existence of some quantities $\mathcal{J}_{\nu\nu'}(t)=-\mathcal{J}_{\nu'\nu}(t)$ satisfying \eqref{ContEq} and \eqref{eq:CurPoi}. One implication was already established in Appendix~\ref{app:SSL2WSL}, we now verify the other by assuming the WSL and providing some possible $\mathcal{J}_{\nu\nu'}(t)$. Number the macro states by their order, $\nu_1 < \nu_2 < \ldots$, and set
\begin{align} 
\mathcal{J}_{\nu_{n+1} \nu_n}(t) 
&= \frac{\mathrm{d}}{\mathrm{d}t} \sum_{\nu\geq \nu_{n+1}} p_\nu(t)\\
\mathcal{J}_{\nu_n \nu_{n+1}}(t) &= -\mathcal{J}_{\nu_{n+1} \nu_n}(t)
\end{align}
and $\mathcal{J}_{\nu \nu'}(t)=0$ otherwise. The WSL in the form \eqref{eq:RMDcond2} then implies \eqref{eq:CurPoi}, while \eqref{ContEq} follows from 
\begin{align}
\frac{\mathrm{d}}{\mathrm{d}t} p_{\nu_n}(t) 
&=\frac{\mathrm{d}}{\mathrm{d}t}  \sum_{\nu\geq \nu_n} p_\nu(t) - \frac{\mathrm{d}}{\mathrm{d}t}  \sum_{\nu\geq \nu_{n+1}} p_\nu(t)\\
&= \mathcal{J}_{\nu_{n} \nu_{n-1}}(t)- \mathcal{J}_{\nu_{n+1} \nu_n}(t)\\
&= \mathcal{J}_{\nu_{n} \nu_{n-1}}(t)+ \mathcal{J}_{\nu_{n} \nu_{n+1}}(t)\\
&= \sum_{\nu'}\mathcal{J}_{\nu_{n} \nu'}(t)\,.
\end{align}

\section{Alternative Quantification of Violations}
\label{app:defdeltaS}

As remarked in Section~\ref{sec:ApproxInc}, an alternative way of quantifying how large the deviations of the SSL are may involve the amount of entropy involved in the violation. That is,
\begin{equation}
	\delta_\ell^S(t) = \int_0^t \mathrm{d}t'  \sum_{\substack{\nu,\nu'\in\mathscr{N}\\\nu< \nu'}} \bigl| S_\qB(\nu)-S_\qB(\nu') \bigr| \: \mathcal{J}^+_{\nu \nu'}(t')
\end{equation}
is the expected total decrease of entropy during $[0,t]$ of a stochastic process $(X_t)$ in $\mathscr{N}$ having distribution $(p_\nu(t))_\nu$ and current $(\mathcal{J}_{\nu\nu'}(t))_{\nu,\nu'}$ for every $t$, and 
\begin{equation}
	\delta_r^S(t) = \int_0^t \mathrm{d}t' \sum_{\substack{\nu,\nu'\in\mathscr{N}\\ \nu > \nu'}} \bigl| S_\qB(\nu)-S_\qB(\nu') \bigr| \: \mathcal{J}^+_{\nu \nu'}(t')
\end{equation}
the expected total increase; the condition we mean is
\be
\delta^S_p(t):=\delta_\ell^S(t)/\delta_r^S(t)\ll 1\,.
\ee

\section{Derivations for Example 1}
\label{app:ex1}

\subsection{Proof of Theorem~\ref{thm:Fourier}}
\label{app:Fourier}

Since $\sqrt{2\pi}|\hat\rho(t)|^2$ is the Fourier transform $\hat{a}(t)$ of the auto-correlation function
\be
a(E) = \int_{\RRR}\mathrm{d}\tilde{E} \, \rho(\tilde{E}) \, \rho(\tilde{E}-E) \,,
\ee
it suffices to show that $(\hat{a})'(t) \leq 0$ for $t>0$, where the prime means derivative.
Note that $a$ is even (substitute $\tilde{E}+E \to \tilde{E}$) and vanishes outside $[-E_{\max},E_{\max}]$, while for $0<E<E_{\max}$
\begin{align}
a(E) &= \int\limits_E^{E_{\max}} \!\! \mathrm{d}\tilde{E} \, \rho(\tilde{E}) \, \rho(\tilde{E}-E) \\
a'(E) &= -\int\limits_E^{E_{\max}} \!\! \mathrm{d}\tilde{E} \, \rho(\tilde{E}) \, \rho'(\tilde{E}-E)  \label{a'}\\
a''(E) &= \int\limits_E^{E_{\max}} \!\! \mathrm{d}\tilde{E} \, \rho(\tilde{E}) \, \rho''(\tilde{E}-E) \label{a''}\\
a'''(E) &= -\rho(E)\rho''(0+)-\!\!\! \int\limits_E^{E_{\max}} \!\! \mathrm{d}\tilde{E} \, \rho(\tilde{E}) \, \rho'''(\tilde{E}-E) \label{a'''}
\end{align}
using the Leibniz integral rule, $\rho(0)=0$ for \eqref{a'} and $\rho'(0)=0$ for \eqref{a''}. Hence, for $0<E<E_{\max}$, 
\be\label{completelymonotonic}
a\geq 0, \quad a'\leq 0, \quad a''\geq 0, \quad a'''\leq 0
\ee
because $\rho$, $\rho'$, $\rho''$, $\rho'''$ are all $\geq 0$ by assumption. Since $\rho,\rho',\rho''$ are continuous and therefore bounded on $[0,E_{\max}]$, both $a'(E)$ and $a''(E)$ tend to 0 as $E\nearrow E_{\max}$, so $a'$ and $a''$ are continuous on $(0,\infty)$ and exist also at $E_{\max}$. Thus, \eqref{completelymonotonic} holds on $(0,\infty)$, except that $a'''$ may not exist at $E_{\max}$ and jump there from $-\rho(E)\rho''(0)$ to 0. Even with this jump, it follows that $a'$ is a concave function on $(0,\infty)$.

A criterion due to Askey \cite{Askey,HS02} says, when specialized to 3 dimensions, that if for an even real function $a$, $-a'$ is non-negative, non-increasing, and convex on $(0,\infty)$, then the function $A(\boldsymbol{x})=a(|\boldsymbol{x}|)$ on $\RRR^3$ is \emph{positive semi-definite}, i.e.,
\be\label{posdefdef}
\sum_{i,j=1}^n c_i^* c_j A(\boldsymbol{x}_i-\boldsymbol{x}_j) \geq 0
\ee
for all $n$, $c_i\in\mathbb{C}$, and $\boldsymbol{x}_i\in\RRR^3$. As we saw above, the conditions are satisfied in our case.

Bochner's theorem \footnote{\url{http://en.wikipedia.org/wiki/Bochner's_theorem}} \cite{RS2} states that if $A\in L^1(\RRR^3)$ is continuous and positive semi-definite, then its Fourier transform is non-negative, $\hat{A}(\boldsymbol{k})\geq 0$ for all $\boldsymbol{k}\in \RRR^3$. In our case, $A$ is continuous because $a$ is, and in $L^1$ because $\int \mathrm{d}E \, E^2 \, a(E) < \infty$ (since $a$ is continuous and compactly supported), so the conditions are satisfied.

A theorem of Grafakos and Teschl \cite{GT13} says, when specialized to 3 dimensions, that whenever $a\in L^1(\RRR)$ is even and $A(\boldsymbol{x})=a(|\boldsymbol{x}|)$ in $L^1(\RRR^3)$, then
\be
\hat{A}(\boldsymbol{k}) = -\frac{1}{|\boldsymbol{k}|} (\hat{a})'(|\boldsymbol{k}|) \quad\text{for }\boldsymbol{k}\neq \boldsymbol{0}\,.
\ee
Thus, $(\hat{a})'(t)\leq 0$ for $t>0$, which is what we wanted to prove.

\subsection{Derivation of Statement~\ref{sta:d1}}
\label{app:d1}

We write $\Hilbert=\Hilbert_\mc$ and choose a purely random orthonormal basis $\varphi_1,\ldots,\varphi_D$ of $\Hilbert$. That is, with the notation $H=\sum_n E_n |\phi_n\rangle \langle \phi_n|$ and $U_{in}:=\langle\varphi_i|\phi_n\rangle$, $U$ is a purely random (Haar-distributed) unitary matrix. Set $\Hilbert_1=\mathrm{span}\{\varphi_1,\ldots,\varphi_{d_1}\}$. Then $\Hilbert_1$ is uniformly distributed among all $d_1$-dimensional subspaces (its ``direction is purely random''), and $P_1=\sum_{i=1}^{d_1} |\varphi_i\rangle \langle \varphi_i|$. 
To obtain a typical $\psi_0\in\SSS(\Hilbert_1)$, we can without loss of generality set $\psi_0:=\varphi_1$. Then
\begin{align}
&\|P_1\psi_t\|^2
= \langle\varphi_1| e^{iHt}P_1 e^{-iHt} |\varphi_1\rangle\\
&= \sum_{n_1,n_2}\langle\varphi_1|\phi_{n_1} \rangle e^{iE_{n_1}t}\langle\phi_{n_1}|P_1|\phi_{n_2}\rangle e^{-iE_{n_2}t} \langle \phi_{n_2} |\varphi_1\rangle\\
&= \sum_{n_1,n_2}\sum_{i=1}^{d_1}U_{1n_1}  e^{iE_{n_1}t}U^*_{in_1}U_{in_2} e^{-iE_{n_2}t} U^*_{1n_2}\,.\label{p1U}
\end{align}
By Weingarten calculus \cite{Collins_23,Wiki}, for any number $n$ of $U$ factors and $U^*$ factors,
\begin{align}
&\EEE\bigl[U^*_{i'_1 j'_1} \cdots U^*_{i'_n j'_n} U_{i_1 j_1} \cdots U_{i_n j_n} \bigr] \nonumber\\
&= \sum_{\sigma,\tau\in S_n} \delta_{i_1i'_{\sigma(1)}} \cdots \delta_{i_n i'_{\sigma(n)}} \delta_{j_1 j'_{\tau(1)}} \cdots \delta_{j_nj'_{\tau(n)}} \mathrm{Wg}(\sigma\tau^{-1},D)\label{Wg}
\end{align}
with $S_n$ the permutation group and Wg the Weingarten function. For $n=2$, the two Weingarten functions are \cite{Wiki}
\begin{align}
\mathrm{Wg}(1^2,D) &= \frac{1}{D^2-1}\\
\mathrm{Wg}(2,D) &= -\frac{1}{D(D^2-1)}\,,
\end{align}
where permutations are denoted by their cycle lengths, so $1^2$ means the identity and $2$ the transposition. Hence,
\begin{align}
&\EEE \bigl[ U_{1n_1}U^*_{in_1}U_{in_2} U^*_{1n_2} \bigr] \nonumber\\
&= 
\delta_{i1}(\mathrm{Wg}(1^2,D)+\mathrm{Wg}(2,D))(1+\delta_{n_1n_2}) \nonumber\\
&+ (1-\delta_{i1})(\mathrm{Wg}(2,D) +\delta_{n_1n_2} \mathrm{Wg}(1^2,D))
\end{align}
So, since $\mathrm{Wg}(1^2,D)+\mathrm{Wg}(2,D) = 1/D(D+1)\to0$ as $D\to\infty$ and $\mathrm{Wg}(2,D)\ll \mathrm{Wg}(1^2,D)$,
\begin{align}
&\EEE\|P_1\psi_t\|^2 \nonumber\\
&= (d_1+1)\bigl(\mathrm{Wg}(1^2,D)+\mathrm{Wg}(2,D)\bigr)\nonumber\\
&+ \Bigl(\mathrm{Wg}(1^2,D)+d_1 \mathrm{Wg}(2,D)\Bigr)\sum_{n_1\neq n_2}e^{i(E_{n_1}-E_{n_2})t}\\
&\approx \frac{1}{D^2} \sum_{n_1\neq n_2}e^{i(E_{n_1}-E_{n_2})t} \label{EEEp1}\\
&\approx \int dE_1 \int dE_2 \, \rho(E_1) \, \rho(E_2) \, e^{i(E_1-E_2)t}\\
&= 2\pi \, |\hat\rho(t)|^2\,,
\end{align}
which is the expression of \eqref{ex1d1}. 

We now show that the variance of $\|P_1\psi_t\|^2$ tends to 0 as $D\to\infty$. We begin by computing the second moment of \eqref{p1U}. Note that Wg depends on the permutation only through its cycle lengths. For $n=4$, all Weingarten functions are listed in \cite{Wiki}; they are all $\mathcal{O}(D^{-p})$ with $p\geq 5$ except when all cycles have length 1 (i.e., for the identity), where $\mathrm{Wg}(1^4,D)=D^{-4}+\mathcal{O}(D^{-6})$. Since the number of summands in \eqref{p1U} is $D^2d_1$, the number of summands in the square of \eqref{p1U} is $D^4d_1^2$, and errors of order $D^{-5}$ can at most add up (for fixed $d_1$) to $\mathcal{O}(D^{-1})$. As a consequence, the second moment of \eqref{p1U} is 
\be\label{leading}
D^{-4}\sum^\sim_{n_1n_2n_3n_4} e^{i(E_{n_1}-E_{n_2}+E_{n_3}-E_{n_4})t} + \mathcal{O}(D^{-1})\,,
\ee
where $\tilde\Sigma$ means that the sum is taken only over pairwise distinct combinations of $n_1,n_2,n_3,n_4$. (Note that the first term is of order $D^0$ because it has of order $D^4$ summands.) Since the square of the  expectation \eqref{EEEp1} of \eqref{p1U} can also be written in the form \eqref{leading}, it follows that $\mathrm{Var}\eqref{p1U}=\mathcal{O}(D^{-1})$ and thus tends to 0 as $D\to\infty$, which completes the derivation of Statement~\ref{sta:d1}.

\section{Derivations for Example 2}

We will now derive the statements about GUE and Wigner matrices; since GUE is a special case of Wigner matrices, we directly give the proof for Wigner matrices. We will first prove Theorem~\ref{thm:1} and then derive Statement~\ref{cor:1} from it.

\subsection{Proof of Theorem~\ref{thm:1}}\label{sec:DerWSLWig}

Using results from Cipolloni et al.\,\cite{CES_22}, we first compute \eqref{WeightsWignerLargeD} and afterwards \eqref{eq:WignerCurrentSSL}. We use the abbreviation 
\be
\varphi_k(t):=J_k(2t)/t\,.
\ee
In \cite{CES_22} it was shown that for a Wigner matrix $H$, a bounded deterministic observable $A$, evolving in time as $A(t)=e^{iH t}A(0)e^{-i H t}$, deterministic vectors $\psi, \phi\in \mathbb{C}^D$ (arbitrary but fixed), and (arbitrarily small) $\xi>0$,
\begin{equation}\label{eq:CipA}
	\langle \psi | A(t) |\phi \rangle 
    =a(t)+ \mathcal{O}(D^{-1/2+\xi} \, t^2)
\end{equation}
for the deterministic approximation
\begin{equation}
    a(t) = \langle \psi | \phi \rangle \frac{\tr A}{D} + \varphi_1^2(t) \: \langle \psi | \left(A-\frac{\tr A}{D}\right) | \phi \rangle
\end{equation}
with very high probability. This means there exist a $C=C(\xi)>0$ and, for every $\varepsilon>0$, a $D_0=D_0(\xi,\varepsilon)$ such that for all $D>D_0$ and $(1-\varepsilon)$-most $H$,
\begin{equation}
    \Bigl| \langle \psi | A(t) |\phi \rangle - a(t) \Bigr| < C \, D^{-1/2+\xi} \, t^2.
\end{equation}
Applying this to our setting, that is $A=P_\nu$, $A(t)=e^{iH t}P_\nu e^{-iH t}$, $\psi=\phi=\psi_0\in \SSS(\Hilbert_{\nu_0})$, and thus $\tr A=\tr P_\nu = d_\nu$, one immediately obtains that for $(1-\varepsilon)$-most $H$, 
\begin{equation}\label{eq:WigWei}
	\Bigl| \W - w_\nu(t) \Bigr| < C \, D^{-1/2+\xi} \, t^2,
\end{equation}
where
\begin{equation}\label{eq:WigWei2}
w_\nu(t) =  \frac{d_\nu}{D} + \varphi_1^2(t) \left( \delta_{\nu \nu_0} - \frac{d_\nu}{D}\right).
\end{equation}
Now
\eqref{WeightsWignerLargeD} follows for $t\ll D^{1/7}$ by using $\xi=1/2-2/7>0$.

Now we compute the current \eqref{eq:WignerCurrentSSL} using \cite[Ex. 2.8]{CES_22}. In the notation used there, we consider the case $k=3$, where $W=H$, $A_1=P_\nu, A_2=P_{\nu'}$, $f_1(H)=e^{iHt}$, $f_2(H)=H$ and $f_3(H)=e^{-iHt}$. As in \cite[Ex. 2.8]{CES_22}, $F_3$ (the deterministic approximation of $e^{iHt}P_\nu H P_{\nu'}e^{-iHt}$) is given by 
\begin{align}
	F_{3} = &P_{\nu}P_{\nu'}\mathrm{sc}_{\circ}[1]\mathrm{sc}_{\circ}[2]\mathrm{sc}_{\circ}[3] \nonumber\\
	&+ D^{-1}\tr( P_{\nu} P_{\nu'}) \mathrm{sc}_{\circ}[1,3]\mathrm{sc}_{\circ}[2] \nonumber\\
	&+ D^{-1}P_{\nu}\tr( P_{\nu'} ) \mathrm{sc}_{\circ}[1]\mathrm{sc}_{\circ}[2,3] \nonumber\\
	&+ D^{-1}P_{\nu'} \tr( P_{\nu} ) \mathrm{sc}_{\circ}[1,2] \mathrm{sc}_{\circ}[3] \nonumber\\
	&+ D^{-1}\tr ( P_{\nu}) \tr( P_{\nu'} ) \mathrm{sc}_{\circ}[1,2,3],
\end{align}
with 
\begin{align}
	\mathrm{sc}_{\circ}[i] &= \langle f_{i} \rangle_{\mathrm{sc}}\label{sccircs1}\\ 
	\mathrm{sc}_{\circ}[i,j] &= \langle f_{i}f_{j} \rangle_{\mathrm{sc}}-\langle f_{i} \rangle_{\mathrm{sc}} \langle f_{j} \rangle_{\mathrm{sc}}\label{sccircs2}\\   
	\mathrm{sc}_{\circ}[1,2,3] &= \langle f_{1}f_{2}f_{3} \rangle_{\mathrm{sc}} -\langle f_{1}f_{2} \rangle_{\mathrm{sc}} \langle f_{3} \rangle_{\mathrm{sc}} \nonumber\\ &- \langle f_{1}f_{3} \rangle_{\mathrm{sc}} \langle f_{2}\rangle_{\mathrm{sc}}
	- \langle f_{2}f_{3} \rangle_{\mathrm{sc}}\langle f_{1} \rangle_{\mathrm{sc}}\nonumber\\
	&+ 2 \langle f_{1} \rangle_{\mathrm{sc}} \langle f_{2} \rangle_{\mathrm{sc}} \langle f_{3} \rangle_{\mathrm{sc}},\label{sccircs3}
\end{align}
where
\begin{equation}
	\langle f \rangle_{\mathrm{sc}} = \int_{-2}^{2} \mathrm{d}x \, f(x) \, \rho_{\mathrm{sc}}(x)\,,
\end{equation}
and $\rho_\text{sc}(x)$ is the semicircle density as in \eqref{eq:WigSemi}.

A simple calculation of \eqref{sccircs1}--\eqref{sccircs3} yields
\begin{align}
	\langle f_{1} \rangle_{\mathrm{sc}} &=\langle f_{3} \rangle_{\mathrm{sc}}=\varphi_1(t), \\
	\langle f_{2}f_{3} \rangle_{\mathrm{sc}} &=-\langle f_{1}f_{2} \rangle_{\mathrm{sc}}=-2i \varphi_2(t),
\end{align}
while $\langle f_{2} \rangle_{\mathrm{sc}} =\langle f_{1}f_{2}f_{3} \rangle_{\mathrm{sc}}=0$ and $\langle f_{1}f_{3} \rangle_{\mathrm{sc}} =1$.
For $\psi_{0}\in\mathbb{S}(\mathscr{H}_{\nu_{0}})$, we find
\begin{equation}
	\langle \psi_{0}\vert F_{3} \vert \psi_{0} \rangle = \frac{2i}{D}\varphi_1(t)\varphi_2(t)(\delta_{\nu'\nu
		_{0}}d_{\nu}-\delta_{\nu\nu_{0}}d_{\nu'}).
\end{equation}
This allows us to state that for $\xi'>0$, $(1-\varepsilon)$-most $H$ with $D>D_0$ are such that 
\begin{equation}\label{eq:CurrApp}
\Bigl| \mathcal{J}_{\nu \nu'}^{\psi_0}(t)- C_{\nu\nu'}^{\psi_0}(t) \Bigr| < C'\, D^{-1/2+\xi'} \, t^3,
\end{equation}
where
\begin{equation}
    C_{\nu\nu'}^{\psi_0}(t) = \frac{4}{D}\varphi_1(t)\varphi_2(t) \bigl( \delta_{\nu'\nu
		_{0}}d_{\nu}-\delta_{\nu\nu_{0}}d_{\nu'} \bigr).
\end{equation}
Now
\eqref{eq:WignerCurrentSSL} follows for $t\ll D^{1/7}$ by using $\xi'=1/2-3/7>0$.

\subsection{Derivation of Statement~\ref{cor:1}}
\label{app:PfCor1}

We compute the cumulative error for the SSL, which then also covers the WSL. According to \eqref{eq:WignerCurrentSSL}, a current only ever occurs between $\nu_0$ and each other $\nu$, so that for $\nu\neq \nu_0$, $\mathrm{d}p_\nu(t)/\mathrm{d}t=\mathcal{J}_{\nu\nu_0}^{\psi_0}(t)$. Note that
$(\mathrm{d}/\mathrm{d}t) \varphi_1^2(t) = - 4\varphi_1(t) \, \varphi_2(t)$. We simply write ``$=$'' instead of ``$\approx$'' in the approximation \eqref{eq:WignerCurrentSSL} because the error $\delta$ can be made arbitrarily small by making $D$ large, so it will not affect the value of $\delta_p$, provided that the number of macro states is not extremely large (say, that $\#\mathscr{N}\ll \sum_{\nu<\nu_0} d_\nu/\delta$) \footnote{For a mathematical proof of Statement~\ref{cor:1}, it would also play a role that $\delta$ is the error for a \emph{single} $t$, while here we would need to estimate the error with which \eqref{eq:WignerCurrentSSL} holds for all $t$ in the relevant interval \emph{simultaneously}.}.

\begin{figure}[t] % Normale figure-Umgebung (einspaltig)
\centering
\includegraphics[width=0.45\textwidth]{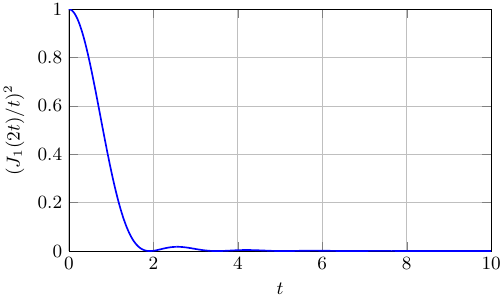}
%\begin{tikzpicture}
%\begin{axis}[
%  domain=0.001:10,
%  samples=500,
%  xlabel={$t$},
%  ylabel={$(J_1(2t)/t)^2$},
%  grid=both,
%  width=\columnwidth,       % Exakt an die Breite einer REVTeX-Spalte angepasst
%  height=0.65\columnwidth,    % Harmonische Hšhe relativ zur Spaltenbreite
%  enlargelimits=false,
%]
%  \addplot[blue, thick] gnuplot {(besj1(2*x) / x)**2};
%   %\addplot[red, thick] gnuplot { (besj1(2*x)/x) *
%  %((besj1(2*x)/x-besj0(2*x))/x)};
%\end{axis}
%\end{tikzpicture}
\caption{\label{fig:Bessel1} The plot shows the function $\varphi_1^2(t)=(J_1(2t)/t)^2$. It is predominantly decreasing, but has oscillations of small amplitude.}
\end{figure}

Thus, at any time $t$ at which $\varphi_1^2(t)$ is decreasing, probability is transported from $\nu_0$ to all other $\nu$'s in the proportion of $d_\nu/D$, and the other way around when $\varphi_1^2(t)$ is increasing. That is,
\begin{align}
	\delta_\ell(t) &= \frac{4}{D} \int_0^t \mathrm{d}t' \sum_{\nu< \nu_0} d_\nu (\varphi_1(t')\varphi_2(t'))^+ \nonumber\\  
    &+ \frac{4}{D} \int_0^t \mathrm{d}t'\sum_{\nu_0 < \nu'} d_{\nu'} (\varphi_1(t')\varphi_2(t'))^{-} \\
		\delta_r(t) &= \frac{4}{D} \int_0^t \mathrm{d}t' \sum_{\nu > \nu_0} d_\nu (\varphi_1(t')\varphi_2(t'))^+  \nonumber\\
        &+ \frac{4}{D} \int_0^t \mathrm{d}t' \sum_{\nu_0 > \nu'} d_{\nu'} (\varphi_1(t')\varphi_2(t'))^{-},
\end{align}
where $x^-=-\min (0,x)$ and $x^+=\max(0,x)$. Since $\nu_0<\eq$ by assumption,
\be
\sum_{\nu<\nu_0} d_\nu < \sum_{\nu<\eq} d_\nu \stackrel{\eqref{deqD1}}{\ll}d_\eq<\sum_{\nu>\nu_0}d_\nu\,,
\ee
so the terms with $\nu<\nu_0$ are negligible compared to those with $\nu>\nu_0$, and
\begin{align}
	\delta_\ell(t) &\approx  \frac{4}{D} \int_0^t \mathrm{d}t'\sum_{\nu_0 < \nu'} d_{\nu'} (\varphi_1(t')\varphi_2(t'))^{-} \\
		\delta_r(t) &\approx \frac{4}{D} \int_0^t \mathrm{d}t' \sum_{\nu > \nu_0} d_\nu (\varphi_1(t')\varphi_2(t'))^+ \,.
\end{align}
That is, the second law is violated whenever $\varphi_1^2(t)$ is increasing, and obeyed whenever it is decreasing. From Figure~\ref{fig:Bessel1}, it is visible that it is predominantly decreasing.

\begin{figure}[t] % Normale figure-Umgebung (einspaltig)
\centering
\includegraphics[width=0.45\textwidth]{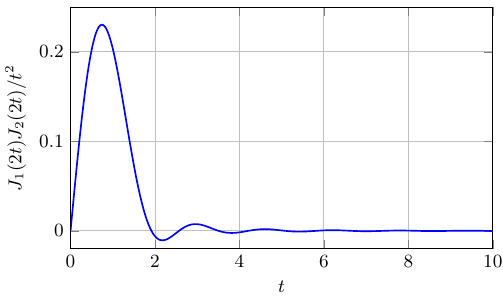}
%\begin{tikzpicture}
%\begin{axis}[
%  domain=0.001:10,
%  samples=500,
%  xlabel={$t$},
%  ylabel={$J_1(2t)J_2(2t)/t^2$},
%  grid=both,
%  width=\columnwidth,       % Exakt an die Breite einer REVTeX-Spalte angepasst
%  height=0.65\columnwidth,    % Harmonische Hšhe relativ zur Spaltenbreite
%  %enlargelimits=false,
%   ymin=-0.02,                % Fixiert die Untergrenze (Gegenschwingungen sichtbar)
%  ymax=0.25,                 % Macht die y-Achse hšher (Peak liegt bei ca. 0.5)
%  enlargelimits=false,
%]
%  \addplot[blue, thick] gnuplot { (besj1(2*x)/x) *
%  ((besj1(2*x)/x-besj0(2*x))/x)};
%\end{axis}
%\end{tikzpicture}
\caption{\label{fig:BesselCur2}The plot shows the function $\varphi_1(t)\varphi_2(t)=J_1(2t)J_2(2t)/t^2$, which is the time-dependent factor in \eqref{eq:WignerCurrentSSL}.}
\end{figure}

Equivalently, the second law is violated whenever (and to the extent that) $\varphi_1(t) \, \varphi_2(t)$ is negative, and obeyed whenever (and to the extent that) it is positive; from Figure~\ref{fig:BesselCur2}, it is visible that the positive contribution is much larger than the negative. We quantify this by computing
\be
        \delta_p(t) = \frac{\delta_\ell(t)}{\delta_r(t)} \approx \frac{\int_0^t \mathrm{d}t' (\varphi_1(t')\varphi_2(t'))^{-}}{\int_0^t \mathrm{d}t'  (\varphi_1(t')\varphi_2(t'))^+ } \,.
\ee

The relevant choice of $t$ (corresponding to the time scale of reaching MATE) is $t\to\infty$ because in the continuum approximation, recurrence times have already been pushed to $\infty$, and approach to MATE coincides with convergence as $t\to\infty$. (As a different way of looking at it, the approximation \eqref{eq:WignerCurrentSSL} becomes exact in the limit $D\to\infty$, in which also the time frame $D^{1/7}$ tends to $\infty$.) One finds that \footnote{These values were obtained using Wolfram|Alpha.}
\begin{align}
	 \int_0^\infty \mathrm{d}t'  (\varphi_1(t')\varphi_2(t'))^+&\approx 0.25638 \\
		 \int_0^\infty \mathrm{d}t' (\varphi_1(t')\varphi_2(t'))^{-}&\approx 0.00639
\end{align}
and therefore 
\be
\lim_{t\rightarrow \infty} \delta_p(t)\approx 0.02492\,. 
\ee
Hence, the WSL and SSL hold within a tolerance of $2.5\%$.

\subsection{Derivation of Remark~\ref{rmk:locdetWig}}
\label{app:locdetWig}

By analogous reasoning as in the proof of Theorem~\ref{thm:1}, we compute $\sigma_{\nu\nu'}$  as given in \eqref{eq:locdetajumprates} and find that, with very high probability,
	\begin{align}
		\tau d_{\nu'}\sigma_{\nu\nu'} &\approx \delta_{\nu\nu'}d_{\nu}\varphi^2_1(\tau)+ \frac{d_\nu d_{\nu'}}{D}-\frac{d_\nu d_{\nu'}}{D}\varphi^2_1(\tau)\,,
\end{align}
while $\sigma_{\nu' \nu}$ can be obtained by exchanging $\nu$ and $\nu'$.
Hence, their ratio is given by
\begin{equation}
\frac{\sigma_{\nu\nu'}}{\sigma_{\nu'\nu}} \approx \frac{\frac{d_\nu}{D}+\varphi^2_1(\tau)(\delta_{\nu\nu'}-\frac{d_\nu}{D})}{\frac{d_{\nu'}}{D}+\varphi^2_1(\tau)(\delta_{\nu\nu'}-\frac{d_{\nu'}}{D})}.
\end{equation}
For $\nu \neq \nu'$ we obtain \eqref{eq:locdetWig}.

\bigskip

\noindent \emph{Acknowledgments.} We thank Cornelia Vogel for the permission to use Figure~\ref{fig:RMDEx} and her and Joscha Henheik for helpful discussions.

\noindent \emph{Use of artificial intelligence.} The proof of Theorem~\ref{thm:Fourier} was suggested to us by ChatGPT version 5.6.

\noindent \emph{Data availability.} Data for Figure~\ref{fig:RMDEx} are available from the authors on  request.

\bibliography{Bibliography.bib}

\end{document}